\documentstyle[aps,prl,twocolumn,psfig]{revtex}
\def\lsim{\lower.8ex\hbox{$\buildrel<\over\sim$}}
\def\gsim{\lower.8ex\hbox{$\buildrel>\over\sim$}}
\begin{document}

\draft
%\twocolumn[\hsize\textwidth\columnwidth\hsize\csname@twocolumnfalse\endcsname
\preprint{\today}
\title{Nucleation Theory for Capillary Condensation}

\author{
Lyd\'eric Bocquet, Fr\'ed\'eric Restagno and Thierry Biben}
\address{Laboratoire de Physique (UMR CNRS 8514)\\
ENS-Lyon, 46 all\'ee d'Italie, 69364 Lyon Cedex 07, France}

\date{\today}

\maketitle

\begin{abstract}
This paper is devoted to the thermally activated 
dynamics of the capillary condensation.
We present a simple model which enables us to identify the critical
nucleus involved in the transition mechanism. This simple model 
is then applied to calculate the nucleation barrier from which 
we can obtain informations on the nucleation time. These results are
compared to the numerical simulation of a Landau-Ginzburg model for
the liquid-vapor interface combined with a Langevin dynamics.

\end{abstract}


\narrowtext
 Humidity is known to strongly affect the mechanical properties
 of many substances, like granular media or porous
 materials \cite{IS85}. Water vapor may for example condense in the pores
 of the medium to form liquid bridges. This phenomenon is the well known capillary
 condensation (see e.g. \cite{Evans}). 
 The Laplace pressure inside such liquid bridges may reach a 
 few atmospheres, and thus results in high adhesion forces
 inside the material.
 More fundamentally, capillary condensation is a
 confinement induced gas-liquid  phase transition. While in the bulk
 the gas phase has a lower free energy, and is thus the stable phase,
 a liquid phase can condense in a pore when the liquid partially wets
 the solid substrate (more specifically when $\gamma_{SL}<\gamma_{SV}$, where $\gamma$
 is a surface tension, $SL$ and $SV$ denote the solid-liquid and the
 solid-vapor interfaces, respectively). A basic model of confinement is
 provided by the slab geometry, for which the fluid is confined
 between two parallel planar solid walls. Macroscopic considerations
 based on this model predict a condensation of the liquid phase below
 a critical distance $H_c$ between the solid surfaces satisfying the condition:
 \begin{equation}
 \Delta\rho~\Delta\mu \simeq {{2(\gamma_{SV}-\gamma_{SL})}/H_c}
 \label{Kelvin}
 \end{equation}
 where $\Delta\rho=\rho_L-\rho_V$ is the difference between the bulk
 densities of the liquid and the gas phase, and $\Delta\mu=\mu_{sat}-\mu$ is
 the (positive) undersaturation in chemical potential ($\mu_{sat}$ is
 the chemical potential at bulk coexistence) \cite{Evans}. The {\em static}
 part of this transition is now well understood and documented, both from
 the experimental \cite{ExpCC,ExpCC2} and theoretical point of view
 \cite{Evans,Evans2}. 

 On the other hand, the {\it dynamics} of the transition have received very
 little attention. Only indirect informations are experimentally
 available in the litterature. Experimental studies of the capillary   
 condensation using the Surface Force Apparatus (SFA) report strong   
 metastability of the gas phase when $H<H_c$, which persists over macroscopic 
 times \cite{ExpCC,ExpCC2}. 
 Furthermore the dynamics have been probed indirectly
 in experiments measuring the time dependent building up of
 cohesion inside divided materials. The latter is indeed a direct measure
 of the dynamical construction of liquid bridges between the grains of the
 medium (see ref. \cite{Boc} for details). From the theoretical side,
 macroscopic arguments for the topologically equivalent drying transition 
 predict extremely large time-scales for evaporation to occur \cite{Yaminski},
 while the latter is found to occur must faster in corresponding lattice-gas 
 simulations \cite{Lum}.

  A detailed theory for the dynamics of the capillary
 condensation providing an estimate of the condensation time is thus 
 still lacking.
 The dynamics of the {\em bulk} gas-liquid
 transition have received on the contrary much more interest \cite{Langer}.
 Away from the spinodal line, in oversaturated situations (when $\mu>\mu_{sat}$),
 a nucleation barrier can be constructed in the bulk situation,
 and a critical nucleus identified.
 The latter takes the form of a spherical droplet, with a radius $R_0\sim 
 \gamma_{LV}/\Delta\mu \Delta\rho$ maximizing the free energy of
 the droplet. Since capillary condensation is also a first order phase
 transition, it should be possible to identify a critical nucleus away
 from the spinodal line \cite{spinodal}. However, the situation is more
 complicated in a confined geometry since the size of the critical nucleus 
 $R_0$ competes with other length scales, like the separation $H$ between
 the walls. 
 In the following, we show how to construct the critical nucleus 
 for capillary condensation. A simplified model 
 keeping only the main ingredients for capillary condensation will be first
 considered.
 The latter has both advantages to allow tractable calculations and to capture 
 the essential features of the involved physics. In our
 case, chemical potential, total volume and temperature are fixed. Our aim is   
 thus
 to find out the saddle-point of the grand potential of the system corresponding
 to the critical nucleus.

\begin{figure}[htbp]
\psfig{file=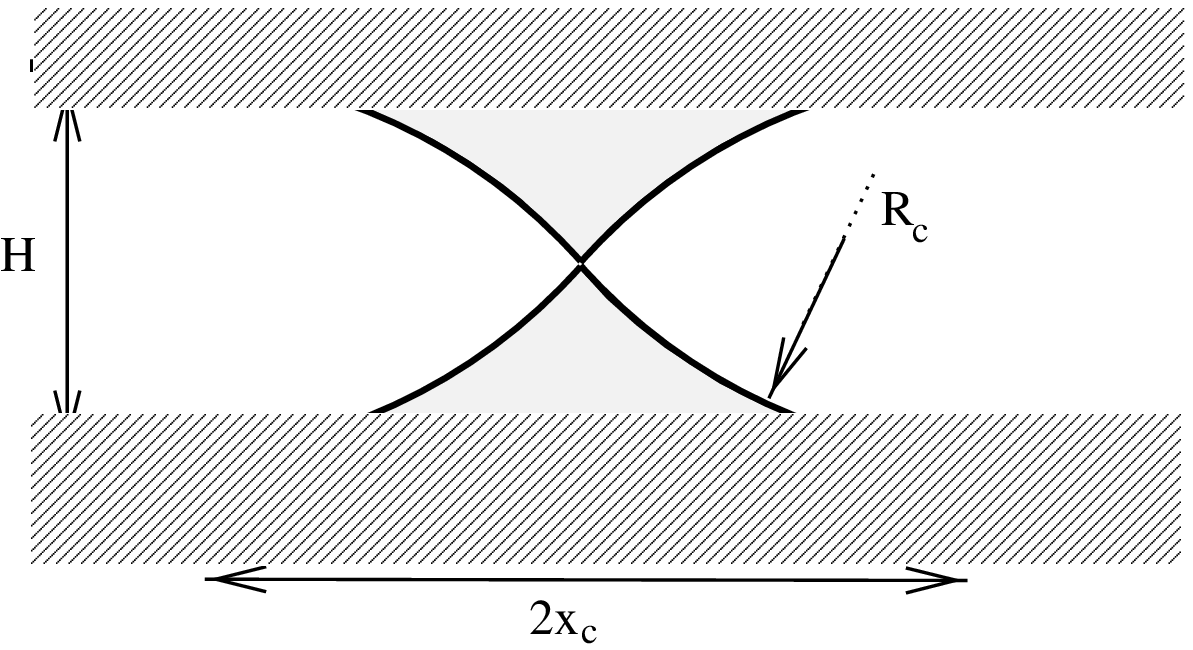,width=7.5cm,height=5.0cm}
\caption{Picture of the critical nucleus for capillary condensation in two dimensions and perfect 
wetting case ($\theta=0$). The radius of curvature of the
meniscus is equal to $R_c=H_c/2$, which is only approximatively verified
within the small slope assumption. See text for details.}
\label{fig1}
\end{figure}

 To simplify further the discussion we first consider a
 two dimensionnal system and a perfect wetting situation:
 $\gamma_{SV}=\gamma_{SL}+\gamma_{LV}$. Both assumptions shall be relaxed
 at the end. In our simplified description, we shall assume that the phase
 equilibrium is determined by macroscopic considerations, so that the $H$
 dependences of the surface tensions are totally neglected. Finally,
 we assume that the system exhibits the mirror symmetry with respect to the $H/2$ plane. 
 Let us consider the situation in which planar liquid films
 of varying thickness $e$ ($e<H/2$) develop on both solid surfaces. 
 Following Evans {\it et al.} \cite{Evans2}, the grand potential of the system may be written
 \begin{equation}
 \Omega=-p_V V_V-p_L V_L+2\gamma_{SL} A+2\gamma_{LV} A
 \label{omega}
 \end{equation}
 where $V_V$ (resp. $V_L$) is the volume of the gas (resp. liquid) and
 $A$ is the surface area.
 Using $V_L=2Ae$, $V_V=A(H-2e)$ and $p_V-p_L\simeq \Delta\rho\Delta\mu$, one gets
 \begin{equation}
 \Delta\omega(e)\equiv{1\over A}(\Omega-\Omega(e=0))=\Delta\rho\Delta\mu~ 2e
 \label{domega}
 \end{equation}
 Note that in the complete wetting situation $\Omega(e=0)$ can be identified
 with $\Omega_V$, the grand-potential of the system filled with the gas
 phase only. The situation $e=H/2$ corresponds to the opposite case where
 the two liquid films merge to fill the pore. The grand potential thus exhibits
 a discontinuity at $e=H/2$ corresponding to the disappearance of the two
 liquid-vapor interfaces, and its value is reduced by $2\gamma_{LV}A$.
 When $e=H/2$, expression (\ref{domega}) must then be replaced by 
 $\Delta\omega(e=H/2)=-\Delta\rho\Delta\mu (H_c-H)$, where $H_c$ is the critical
 distance defined in eq. (\ref{Kelvin}).  One may note
 that the minimum of the grand potential corresponds to a complete filling
 of the pore by the liquid phase when $H<Hc$, as expected.
 Up to now, liquid-vapor interfaces were assumed to be planar. If we
 allow deformation of the interfaces, {\it i.e.} the thickness $e$ is now a
 function of the lateral coordinates, the corresponding cost has to be added
 to the grand potential. Due to the mirror symmetry assumption, the two films 
 are identical, so that one finds eventually
 \begin{equation}
 \Delta\Omega_{tot}=\int dx~\left\{{\gamma_{LV}}\vert\nabla e\vert^2 + \Delta \omega(e)\right\}
 \label{eq1}
 \end{equation}
 with $\Delta\Omega_{tot}=\Omega(\{e\})-\Omega_V$ and a small slope hypothesis
 has been made. Extremalization of the grand potential in two dimensions
 leads to the following Euler-Lagrange equation for $e(x)$, where $x$ denote
 the lateral coordinate:
 \begin{equation}
 2\gamma_{LV} {d^2e\over{dx^2}} - {d\Delta \omega(e)\over de} =0
 \label{eq2}
 \end{equation}
 This last equation is formally equivalent
 to the mechanical motion of a particle of mass $2 \gamma_{LV}$, with position 
 $e$ in the external
 potential $-\Delta \omega(e)$; $x$ plays the role of time. We look
 for solutions satisfying $e=0$ and
 $de/dx=0$ at infinity. Starting from $e=0$, the ``particle'' is uniformly 
 accelerated until it reaches $e=H/2$. We can choose this last point to
 fix the origin i.e. $e(x=0)=H/2$. At this point, the discontinuity in the potential
 induces a specular reflexion, similar to a collision of the particle
 with a hard wall : $de/dx$ is therefore discontinuous and antisymetric at $x=0$.
 After a straightforward calculation the complete solution, depicted in fig
 \ref{fig1}, can now be obtained and is found to have a spatial extension
 $x_c=\sqrt{H R_c}$, with $R_c=H_c/2$. Explicitly one gets 
%\begin{equation}
% e(x)={(x-x_c)^2\over {2R_c}} \hbox{    for $x\in [0;x_c]$}
% \label{eq3}
% \end{equation}
$e(x)={(|x|-x_c)^2/{2R_c}}$ for $x\in [-x_c;x_c]$ and zero otherwise.
 Let us note that the cusp in the solution at $x=0$ stems from 
 the discontinuity of $\Delta\omega$ at $e=H/2$ resulting from the assumption
 of an infinitesimely narrow liquid-vapor interface.
 Condensation thus occurs through the
 exitation of short wavelength fluctuations, in agreement with the simulations
 results for the drying transition \cite{Lum}.
 The corresponding energy of the nucleus (per unit length in the perpendicular direction) 
 can be calculated by integrating eq. (\ref{eq1}),
 to obtain
 \begin{equation}
 \Delta \Omega^{\dag} = {4\over 3} (\Delta\mu \Delta\rho \gamma_{LV})^{1/2}
 H^{3/2}
 \label{eq4}
 \end{equation}
 This energy can be identified with the energy barrier to overcome in order
 to condense the liquid phase from the metastable gas phase. When  
 the energy barrier is not too small (compared to $k_BT$), the time needed 
 to condense can be estimated to be $\tau =\tau_0 \exp(\Delta\Omega^{\dag} 
 /k_BT)$, with $\tau_0$ a ``microscopic'' time.
 It is easy to check that $\Delta \Omega^{\dag} $ corresponds to a saddle-point of
 the grand-potential. It is greater than both free energies of the gas and
 liquid phases. Moreover $\Delta \Omega^{\dag} $ is smaller than the 
 free energy of any other configuration maximizing the grand potential 
 since it is the only solution of finite extension. 
 We postpone the physical interpretation of the results to the end of the
 letter. We just point out that the parabolic solution obtained above
 is the small slope approximation to the circle with radius of curvature $R_c$.

 In order to verify the previous results, we have conducted numerical simulations
 of the capillary condensation in two dimensions. We start with a 
 Landau-Ginzburg model for the grand potential of the system confined between
 two wall. In terms of the local density $\rho({\bf r})$, we write the ``excess''
 part of the grand potential $\Omega^{ex}=\Omega+P_{sat}V$, where $P_{sat}$ is the pressure
 of the system at coexistence, as
 \begin{equation}
 \Omega^{ex}= \int d{\bf r}\left\{ {m\over 2}  \vert\nabla \rho
 \vert^2 + W(\rho) + \left(\Delta\mu+V_{ext}(z)\right)\rho \right\}
 \label{eq5}
 \end{equation}
 In this equation, $m$ is a phenomenological parameter; $V_{ext}(z)$ is the confining external potential,
 which we took for each wall as $V_{ext}(z)=-\epsilon (\sigma/(\Delta 
 z+\sigma))^3$, with $\Delta z$ the distance to the corresponding wall;
 $\epsilon$ and $\sigma$ have the dimensions of an energy and a distance.
 $W(\rho)$ can be interpreted as the negative of the 
 ``excess'' pressure $\mu_{sat}\rho-f(\rho)-P_{sat}$, with $f(\rho)$ the free-energy
 density \cite{RW}. As usually done, we assume a phenomenological double well
 form for $W(\rho)$ : $W(\rho)=a(\rho-\rho_V)^2(\rho-\rho_L)^2$, where $a$ is a
 phenomenological parameter \cite{Safran}. The system
 is then driven by a non-conserved Langevin equation for $\rho$~:
 \begin{equation}
 {\partial\rho \over{\partial t}}= -\Gamma {\delta \Omega^{ex} \over {\delta\rho}} + \eta({\bf r},t)
 \label{langevin}
 \end{equation}
 where $\Gamma$ is a phenomenological friction coefficient and $\eta$ is 
 a Gaussian noise field related to $\Gamma$ through the fluctuation-dissipation 
 relationship \cite{Dynamics}.
 %\begin{equation}
 %\langle\eta({\bf r},t)\eta({\bf r}^\prime,t^\prime)\rangle
 %=2\Gamma k_BT \delta({\bf r}-{\bf r}^\prime) \delta(t-t^\prime)
 %\label{FlucDiss}
 %\end{equation}
 An equivalent model has been  successfully used for the (bulk) classic nucleation
 problem \cite{Valls}. Physically, the non-conserved
 dynamics assume an infinitely fast transport of matter in the system, which is
 justified in view of the time-scale involved for condensation.
 We solved (\ref{langevin}) by numerical integration using standard methods,
 identical to those of ref. \cite{Valls}. The units of energy, length 
 are such that $\sigma=\epsilon =1$. Time is in unit of $t_0=(\Gamma \epsilon\sigma^2)^{-1}$ 
 with $\Gamma={1\over 3}$. In these units, 
 we took $m=1.66$, $a=3.33$, $\rho_L=1$, $\rho_V=0.1$. Typical values of the chemical 
 potential and temperature are $\Delta\mu\sim 0.016$, $T\sim 0.06$ (which is 
 roughly half the energy barrier between vapor and liquid with the form
 for $W(\rho)$ used in our model). Periodic boundary condition with periodicity 
 $L_x$ were applied in the lateral direction. 
\begin{figure}[htbp]
\psfig{file=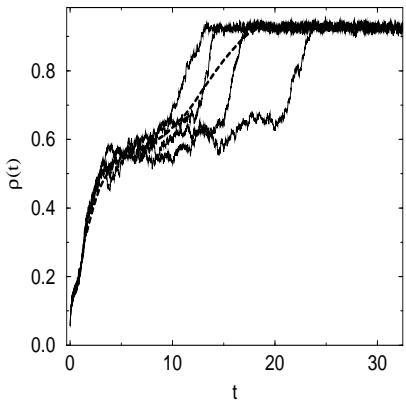,width=6cm,height=6.cm}
\caption{Averaged density as a function of time $t$ (in units of $t_0$) for a few realizations
of the noise ($H=13$, $\Delta\mu=0.016$ and $T=0.07$). The dashed
line is the average over all the realizations, $\bar\rho(t)$.}
\label{fig2}
\end{figure}
Typically $L_x\sim 2H$ was used,
 but we have checked that increasing $L_x$ up to $20 H$ does not affect the results 
 for the activation dynamics. We emphasize that this lack of sensitivity is not an
 obvious result since it is known that the amplitude of capillary waves increases with
 the lateral dimension of the system for {\it free interfaces} \cite{Evans}. In our case
 however, the long-range effects of the fluctuations of the liquid film are expected to be
 screened due to the presence of the external potential.  Moreover, as predicted
 by the model, nucleation should occur via the excitation of localized fluctuations.
 The observed insensitivity of the results with respect to finite size effects is
 then an encouraging feature for the model presented in this letter.

\begin{figure}[htbp]
\psfig{file=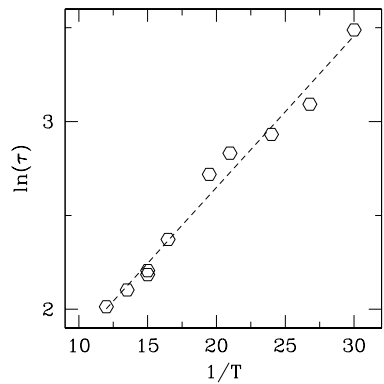,width=6cm,height=6cm}
\caption{Logarithm of condensation time as a function of the inverse 
 temperature 
($\Delta\mu=0.016$, $H=13$). The dashed line is least-square fit of the datas.}
\label{figT}
\end{figure}

\begin{figure}[htbp]
\psfig{file=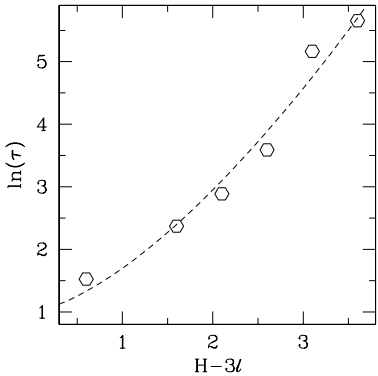,width=6cm,height=6cm}
\caption{Logarithm of condensation time as a function of the ``effective'' 
width of the slab $H-3\ell$ for fixed $\Delta\mu=0.016$. The dashed line is the theoretical
prediction $\ln(\tau)=\ln(\tau_0)+\alpha (H-3\ell)^{3/2}$. The two parameters $\ln(\tau_0)$ and 
$\alpha$ have been obtained from a least-square fit of the datas in a $\ln(\tau)$ versus $(H-3\ell)^{3/2}$ plot.}
\label{fig3}
\end{figure}

 The simulated system is initially a gas state filling the whole pore, and its
 evolution is described by eq. (\ref{langevin}). A typical evolution of the mean density
 in the slit $\rho(t)$ is plotted on fig \ref{fig2}. An average over different 
 realizations (from 10 to 30)
 is next performed to get an averaged time-dependent density $\bar{\rho}(t)$.
 As expected \cite{Evans2}, due to the long range nature of the external
 potential a thick liquid film of thickness $\ell$ 
 rapidly forms on both walls on a short
 time scale $\tau_1$ ($\ell\simeq 3.8 \sigma$ and $\tau_1\approx 5 t_0$ in our case). 
 The first step
 of the dynamics is thus the wetting of the solid substrates. This process is 
 not thermally activated. In a second stage,
 fluctuations of the interfaces around their mean value $\ell$ induce after 
 a while a sudden  coalescence of the films 
 (see fig. \ref{fig2}). This second process has a characteristic time $\tau$.
 It is numerically convenient to define 
 the total coalescence time, $\tau_1+\tau$, as the time 
 for the average density  in the slab between the two wetting films to
 reach $(\rho_V+\rho_L)/2$ \cite{Valls}, which corresponds 
 in our case to the condition $\bar{\rho}(\tau+\tau_1)\approx 0.8$.  
 The physical 
 results do not depend anyway on the precise definition of $\tau$.
 In fig. \ref{figT}, we plot the variation of $\ln(\tau)$ as a function of
 the inverse temperature $1/T$.
 As expected, far from the spinodal ({\it i.e.}, for large enough $H$, $H\gsim
 3{\ell}$), $\tau$ is found to obey an Arrhenius law $\tau=\tau_0\exp(\Delta 
 \Omega^{\dag}/k_BT)$, where $\Delta \Omega^{\dag}$ is
 identified as the energy barrier for nucleation. We now focus on the variations
 of $\ln(\tau)$ as a function of  $L_x$ and $H$,
 which one assumes to be mainly controlled by the variation of
 $\Delta \Omega^{\dag}$. First, as already noticed above, we found no variation of 
 $\ln(\tau)$ as a function of $L_x$, in agreement with our prediction of
 a localized critical nucleus. The dependence on $H$ ($\Delta\mu$ being fixed)
 is plotted on fig. \ref{fig3}. The previous model predicts a $H^{3/2}$ 
 dependence. However the long range of the external potential (of the 
 van der Waals type), produces thick wetting films whose thickness
 has to be substracted from $H$. A more careful analysis of the critical nucleus
 shows in fact that 
 the total effective thickness of the films has to be replaced by
 $3{\ell}$ (instead of $2{\ell}$), in 
 order to take correctly into account the long range character of the external 
 potential. This result is in agreement with other theoretical
 and experimental findings for capillary condensation in the presence of
 van der Waals forces \cite{Evans2,Jerome}. As seen on fig. \ref{fig3},
 a good agreement with the theoretical prediction is found. Dependence on the
 other parameters ($\Delta\mu$, $\epsilon$, ...) shall be discussed in a longer
 version of this paper.

 The simulation results in the 2D perfect wetting case are thus in agreement 
 with the theoretical predictions.  Now it is possible to relax the assumptions
 done in the presentation above, {\it i.e.} perfect wetting, two dimensionnal
 system, small slopes. More generally, one may realize that the
 extremalization of the grand potential leads in one hand to the usual Laplace
 equation, relating the local curvature $\kappa$ to the pressure drop
 ${\gamma_{LV} \kappa}=\Delta p\simeq \Delta \mu\Delta \rho$; and in the
 other hand, it fixes the contact angle of the meniscus on the solid substrate
 according to Young's law $\gamma_{LV} \cos\theta=\gamma_{SV}-\gamma_{SL}$.
 In our case however, the corresponding nucleus corresponds to a maximum
 of the grand potential.
 In two
 dimension the general solution is of the same geometrical form as the
 one obtained above and pictured in fig \ref{fig1}, with a corresponding
 nucleation energy
 \begin{equation}
 \Delta\Omega^{\dag}=\gamma_{LV} H_c \left\{ {\alpha}
 -{\sin \alpha} \cos(\alpha+2\theta)\right\}
 \label{total}
 \end{equation}
 with $\alpha$ defined through $\cos(\alpha+\theta)=\cos(\theta)-H/H_c$.
 The previous result, eq. (\ref{eq4}), is recovered in the limiting case
 $\theta=0$, $H\ll H_c$.
 In three dimensions, the nucleus takes the form of a liquid
 bridge of finite lateral extension connecting the two solid surfaces, due
 in particular to the supplementary (negative) axisymmetric curvature. 
 When $H$ is small compared to $H_c$, this predicts $\Delta \Omega^{\dag}
 \propto \gamma_{LV} H^2 + {\cal O}(\Delta\mu)$, but the full
 dependence on $\Delta\mu$ for any $H$ needs a numerical resolution. This will
 be done
 in a forthcoming paper together with the corresponding simulations.

\thanks
The authors would like to thank E. Charlaix, J. Crassous and J.C. Geminard
for many interesting discussions. This work has been partly supported by
the PSMN at ENS-Lyon, and the MENRT under contract 98B0316.


\begin{references}
\bibitem{spinodal} For sufficiently small $H$, it can be shown that the liquid films coating the solid surfaces become unstable due to fluid-fluid 
interactions and grow to fill the slab (see ref. \cite{ExpCC2} for details).
\end{references}
\end{document}